\documentclass[%
 reprint,
superscriptaddress,
nofootinbib,
 showframe,
 amsmath,amssymb,
 aps,
 prd,
]{revtex4-2}

\usepackage{graphicx}
\usepackage{dcolumn}
\usepackage{bm}
\usepackage{hyperref}
\usepackage{xcolor}

\begin{document}

\preprint{APS/123-QED}

\title{Neural network based emulation of galaxy power spectrum covariances - A reanalysis of BOSS DR12 data}

\author{Joseph Adamo}
    \email{jadamo@arizona.edu}
    \affiliation{Department of Astronomy/Steward Observatory, University of Arizona, 933 North Cherry Avenue, Tucson, AZ 85721, USA\\}
\author{Hung-Jin Huang}
    \affiliation{Department of Astronomy/Steward Observatory, University of Arizona, 933 North Cherry Avenue, Tucson, AZ 85721, USA\\}
\author{Tim Eifler}
    \affiliation{Department of Astronomy/Steward Observatory, University of Arizona, 933 North Cherry Avenue, Tucson, AZ 85721, USA\\}
    \affiliation{Department of Physics, University of Arizona, 1118 E Fourth Str, Tucson, AZ 85721, USA\\}

\date{\today}

\begin{abstract}
We train neural networks to quickly generate redshift-space galaxy power spectrum covariances from a given parameter set (cosmology and galaxy bias). This covariance emulator utilizes a combination of traditional fully-connected network layers and transformer architecture to accurately predict covariance matrices for the high redshift, north galactic cap sample of the BOSS DR12 galaxy catalog. We run simulated likelihood analyses with emulated and brute-force computed covariances, and we quantify the network's performance via two different metrics: 1) difference in $\chi^2$ and 2) likelihood contours for simulated BOSS DR 12 analyses. We find that the emulator returns excellent results over a large parameter range. We then use our emulator to perform a re-analysis of the BOSS HighZ NGC galaxy power spectrum, and find that varying covariance with cosmology along with the model vector produces $\Omega_m = 0.276^{+0.013}_{-0.015}$, $H_0 = 70.2\pm 1.9$ km/s/Mpc, and $\sigma_8 = 0.674^{+0.058}_{-0.077}$. These constraints represent an average $0.46\sigma$ shift in best-fit values and a $5\%$ increase in constraining power compared to fixing the covariance matrix ($\Omega_m = 0.293\pm 0.017$, $H_0 = 70.3\pm 2.0$ km/s/Mpc, $\sigma_8 = 0.702^{+0.063}_{-0.075}$). This work demonstrates that emulators for more complex cosmological quantities than second-order statistics can be trained over a wide parameter range at sufficiently high accuracy to be implemented in realistic likelihood analyses. 
\end{abstract}

\maketitle

\section{\label{sec:intro}Introduction}

The amount of data available to cosmologists has greatly expanded in recent years.  Spectroscopic galaxy surveys like the Baryon Oscillation Spectroscopic Survey (BOSS, \cite{BOSS, SDSS-IV}), photometric experiments like the Kilo-Degree Survey \cite{KiDS}, the Dark Energy Survey (DES, \cite{DES-Y3-results}), and the Hyper-Surpime Camera (HSC, \cite{HSC-Y3}), and CMB measurements (e.g. Planck, \cite{Planck-2018}), have provided exquisite data for constraining our cosmological model. This trend will only continue with current and upcoming experiments like the Dark Energy Spectroscopic Instrument (DESI, \cite{DESI}), Euclid,\footnote{\url{https://www.cosmos.esa.int/web/euclid}} Rubin Observatory's Legacy Survey of Space and Time\footnote{\url{https://rubinobservatory.org/}} (LSST, \cite{LSST}), the SPHEREx explorer mission \cite{SPHEREx}, and the Nancy Grace Roman space telescope.\footnote{\url{https://roman.gsfc.nasa.gov/}} As this trajectory towards more and higher quality data continues, our analysis methods must improve to accurately extract cosmological information.

In typical analyses of cosmological datasets, the covariance matrix associated with the measured data vector determines the final uncertainty on inferred parameters \cite{Krause-2017-cov, DES-Y3-covariance}. Methods for producing covariance matrices fall into one of three categories: estimating the covariance using the data directly, estimating the covariance using numerical simulations, and computing the covariance matrix from an analytical expression. 

In the first approach, one partitions the data via bootstrapping or jackknifing and recombines the samples to obtain a noisy estimate of the covariance. The resulting estimate does not depend on the choice of model or assumed cosmology. The second approach involves estimating the covariance using mock data from simulations. This approach has been used in real-space analyses with HSC data \cite{Shirasaki-2017-MocksVSJK,Hamana-2019-HSCtpcf}, as well as for full-shape analyses of BOSS DR12 power spectra \cite{BOSS}. Using simulations improves the accuracy of covariances on nonlinear scales, where analytical calculations fail. However, this approach typically requires several thousand realizations to minimize sampling noise, representing a significant computational cost \cite{Taylor-2013-Precision,Dodelson-2013}. Since both of these approaches estimate the covariance matrix from a finite number of realizations, the covariance estimation is noisy and hence the analysis technically requires the use of a multivariate student-t distribution as the likelihood function instead of a multivariate Gaussian \cite{Sellentin-2016-CovStudentsT}. 

The third approach computes the covariance matrix from an analytical model. This approach was implemented in the DES science verification, Year 1 and Year 3 phases \cite{Krause-2017-Cosmolike,Krause-2017-cov,Krause-2021,DES-Y3-covariance}, the HSC analysis in Fourier space \cite{Hikage-2019-HSC}, newer BOSS analysis of RSD power spectra \cite{Wadekar-2020}, and cosmological analyses from the KiDS collaboration \cite{KiDS,Asgari-2021,Heymans-2021}. This method is also used in the context of cluster cosmology using weak lensing \cite{Gruen-2015,McClintock-2019-DESY1RMWL,Murata-2019}, where one can straightforwardly compute covariances using the halo model. This method is computationally less expensive than other methods, but has the drawback that accuracy breaks down at nonlinear scales. Recent studies have worked to combine analytic and simulation-based methods \cite{Friedrich-2016-CovEst,Wu-2019-ClusterWLCovs, Friedrich-Eifler-2017}. These hybrid approaches provide accurate covariance estimates at different physical scales while also reducing the number of simulation realizations required.

Analytic and simulation-based covariance calculations require a set of input parameters, in particular cosmological parameters, that are a priori unknown since they are the target of the inference process. As a consequence, it has been suggested in the literature (e.g. Refs, \cite{Eifler-2009, Labatie-2012, Kodwani-2018-covs, Jee-2013}) to update the covariance matrix at each point of an MCMC chain, similar to the model vector. Such a method can affect the resulting parameter contours when compared to keeping the covariance fixed. As a recent example, the Euclid collaboration found noticeable improvements to constraining power and the FoM metric when varying the covariance matrix in a simulated cluster correlation function analysis \cite{Euclid-Covariance}. 

While conceptually straightforward, this technique runs into several problems. As explained in Ref. \cite{Carron-2013}, a parameter dependent (Gaussian) covariance violates the Cram\'{e}r-Rao bound when assuming that the underlying field is Gaussian, and when approximating the likelihood function of the field's second-order statistic as a multivariate Gaussian. The correct likelihood function for second-order statistics -- a Wishart distribution which asymptotically converges to a Gaussian when there are many modes per bin -- does not depend on a covariance matrix. It correctly assumes that for a Gaussian random field, all information is already contained in the mean of the second-order statistic. In contrast, the Gaussian distribution assumes the mean and variance of our estimator are uncorrelated, which causes the covariance to contribute information. 

The above argument holds for the case of an underlying Gaussian field. However, when pushing to smaller scales in weak lensing or galaxy clustering, the field becomes non-Gaussian and higher order moments carry information through additional covariance terms. Ideally, one would construct a joint likelihood for varying both the mean and covariance matrix in this case, however such an approach is beyond the scope of this paper. Here, we set out to address the major computational cost of computing these objects at different cosmologies.

Updating the covariance at every step of an analysis is expensive in the case of analytical models already, and simply non-feasible for simulation based techniques. In this paper, we present a fast method for emulating covariance matrices using machine learning. We develop a deep neural network to directly emulate covariance matrices by combining traditional fully-connected network frameworks with transformer architectures \cite{Attention-Paper}. This emulator is trained to emulate BOSS redshift-space galaxy power spectrum multipole covariance matrices calculated using perturbation theory. We then use our emulator to run a BOSS analysis varying the covariance with cosmology and compare the resulting contours to an analysis in which the covariance is kept fixed. We use the BOSS DR12 data vector as presented in Ref. \cite{Beutler-Data}, which has been previously analyzed in several works \cite{Chen22, Kobayashi22, Philcox_Ivanov22}.

This paper will proceed as follows. Section \ref{sec:theory} covers the covariance matrix model we emulate, as well as the power spectrum model it is built on. Section \ref{sec:net} describes the network architecture and how we train our emulator. Section \ref{sec:tests} describes and presents the performance metrics and simulated likelihood analyses used to determine emulator accuracy. Finally, we use our emulator to re-analyze BOSS data with cosmology-dependent covariance in Section \ref{sec:BOSS}, and summarize our findings in Section \ref{sec:conclusion}.

\section{Theoretical Background}
\label{sec:theory}

\subsection{Power Spectrum Model}
\label{subsec:Pk}

This section largely follows the model description in Section 3.1 of Ref. \cite{Ivanov-2020}. Our galaxy power spectrum model uses 1-loop Eulerian perturbation theory (PT), which works by splitting the power spectrum multipole prediction into four distinct pieces,

\begin{equation}
    P_{l}(k) = P_l^\text{tree}(k) + P_l^\text{1 loop}(k) + P_l^\text{noise}(k) + P_l^\text{ctr}(k) \,.
\label{eq:pt-model}
\end{equation}

We are suppressing time-dependence by assuming all terms are calculated at the effective redshift of our data sample. The first term in Eq. \ref{eq:pt-model} is the tree-level contribution to the power spectrum given by the Kaiser formula \cite{Kaiser-1987},

\begin{equation}
    P_g(k, \mu) = (b_1 + f \mu^2)^2 P_{lin}(k),
\label{eq:kaiser}
\end{equation}
where $f$ is the growth factor and $b_1$ is the scale-independent linear bias. Next, $P_l^\text{1 loop}(k)$ corrects for nonlinear effects from gravitational evolution and redshift-space distortions. For further detail on these 1-loop corrections, we refer the reader to Refs. \cite{CLASS-PT, Fonseca-2020, Perko-2016}. The full expression for this term is found by expanding the nonlinear galaxy density contrast using the basis

\begin{equation}
    \delta_g = b_1 \delta + \frac{b_2}{2} \delta^2 + b_{\mathcal{G}_2} \mathcal{G}_2 \,,
\end{equation}

where $\delta$ is the nonlinear matter density contrast, $\mathcal{G}_2$ is the tidal field operator in Fourier space, and both $b_2$ and $b_{\mathcal{G}_2}$ are additional bias parameters. Next, $P_l^\text{noise}(k)$ represents stochastic contributions to the power spectrum, which we model with a uniform shot noise parameter ($P_0^\text{noise}(k) = P_\text{shot}$, $P_2^\text{noise}(k) = 0$). 

Lastly, $P_l^\text{ctr}(k)$ encompasses the counterterms, which corrects for spurious dependence on small-scale physics in the 1-loop term \cite{Senatore-2014}. Following Ref. \cite{CLASS-PT}, we apply two multipole-specific counterterms (with free parameters $c_0$ and $c_2$) and one next-to-leading order counterterm (with free parameter $\tilde{c}$) applied to both multipoles. Each counterterm attempts to correct for different effects, and their scale-dependence are set by symmetry arguments \cite{Senatore-2014, Perko-2016}.

To summarize, our power spectrum model is based on 1-loop perturbation theory and contains seven nuisance parameters ($b_1, b_2, b_{\mathcal{G}_2}, c_0^2, c_2^2, \tilde{c}, P_\text{shot}$). When adding IR resummation and the Alcock-Paczynski effect, this model produces accurate predictions up to $k = 0.3$ h/Mpc.

\subsection{Covariance Model}
\label{subsec:covariance}

Similar to the power spectrum, we utilize a PT-based model to compute covariance matrices. For further details, we refer the reader to Ref. \cite{CovaPT}. In this scheme, the covariance matrix is represented as a combination of Gaussian, regular trispectrum, and super-sample covariance (SSC) terms,

\begin{equation}
    C = C_G + C_{T0} + C_{BC} + C_{LA} \,.
\label{eq:cov_model}
\end{equation}

The Gaussian covariance term includes contributions from the disconnected four-point function, and is explicitly written as follows \cite{CovaPT},

\begin{equation}
    C_{l_1, l_2}(k_1, k_2) = \sum_{l_1', l_2'} P_{l_1'}(k_1) P_{l_2'}(k_2) \mathcal{W}_{l_1, l_2, l_1', l_2'}(k_1, k_2)\,,
\end{equation}

where $P_l (k)$ are the predicted power spectrum multipoles from Section \ref{subsec:Pk}, and $\mathcal{W}$ is the window kernel, which includes effects from the survey geometry and the changing line-of-sight direction. $C_G$ is computationally cheap to compute since it is essentially the product of two power spectra. The cosmology and bias parameter dependence of $C_G$ enter through the power spectrum.

The remaining three terms in Eq. \eqref{eq:cov_model} depend on the connected four-point function and are commonly known as non-Gaussian covariance. The first of these terms ($C_{T0}$) is found by computing the regular trispectrum, which assumes either an infinite survey volume or periodic boundary conditions. We calculate the trispectrum at the tree level, which is formally the same order as the 1-loop power spectrum and makes $C_{T0}$ the most expensive covariance term to calculate. We set all third-order bias parameters to $0$ during our calculations, meaning this term depends on the same parameters as our power spectrum model.

Next, we take finite survey effects into account with $C_{BC}$ and $C_{LA}$. The first term corrects for the coupling of long-wavelength modes that are not covered by the survey volume, an effect commonly known as beat coupling \cite{Takada-2013}. The second term accounts for differences between the local and true average galaxy densities, which again depends on the survey volume \cite{Roland-de-Putter-2012}. Both $C_{BC}$ and $C_{LA}$ together make up the super-sample covariance (SSC), which has been studied extensively in the context of galaxy surveys \cite{Li-2014, Mohammed-2016}. Similar to $C_G$, these terms are computationally cheap to compute after survey-specific window kernels are calculated.

To summarize, our covariance matrix model is composed of a disconnected Gaussian term $C_G$, a connected trispectrum term calculated at tree-level $C_{T0}$, and a SSC term composed of beat-coupling and local average affects $C_{BC} + C_{LA}$. The model depends on the same parameters as our power spectrum model, either indirectly through $C_G$, or directly through calculations of the regular trispectrum $C_{T0}$.

\section{\label{sec:net}Emulating Covariance Matrices}

\subsection{Training Dataset}

Our training set consists of analytic covariance matrices calculated using the model defined in Section \ref{subsec:covariance}. We use the public code \verb|CovaPT|\footnote{\url{https://github.com/JayWadekar/CovaPT}} to calculate monopole and quadropole galaxy power spectrum covariance for the north galactic cap CMASS galaxy sample of the BOSS DR12 catalog. Our choice of k binning follows Ref. \cite{Wadekar-2020}, so $k_\text{max} = 0.25$ h/Mpc and $\Delta k = 0.01\ h/Mpc$. We generate a Latin hypercube of $745000$ cosmology and bias parameters to pass to \verb|CovaPT|, with the specific parameters and ranges given in Table \ref{table:cosmology}.

For each covariance matrix in our training set, we compute the lower Cholesky decomposition $C = L L^T$, which is guaranteed to exist for positive-definite matrices. This process results in the lower-triangular matrix $L$, which encapsulates all the independent information present in the covariance. To reduce the dynamic range of values to emulate, we then take the element-wise logarithm, which normalizes the training set in a reversible fashion. Our emulator is thus trained to reproduce $\log L$, which can quickly be transformed to a full covariance matrix.

\subsection{\label{subsec:architecture}Neural Network Architecture}

\begin{figure*}
    \centering
    \includegraphics[width=\linewidth]{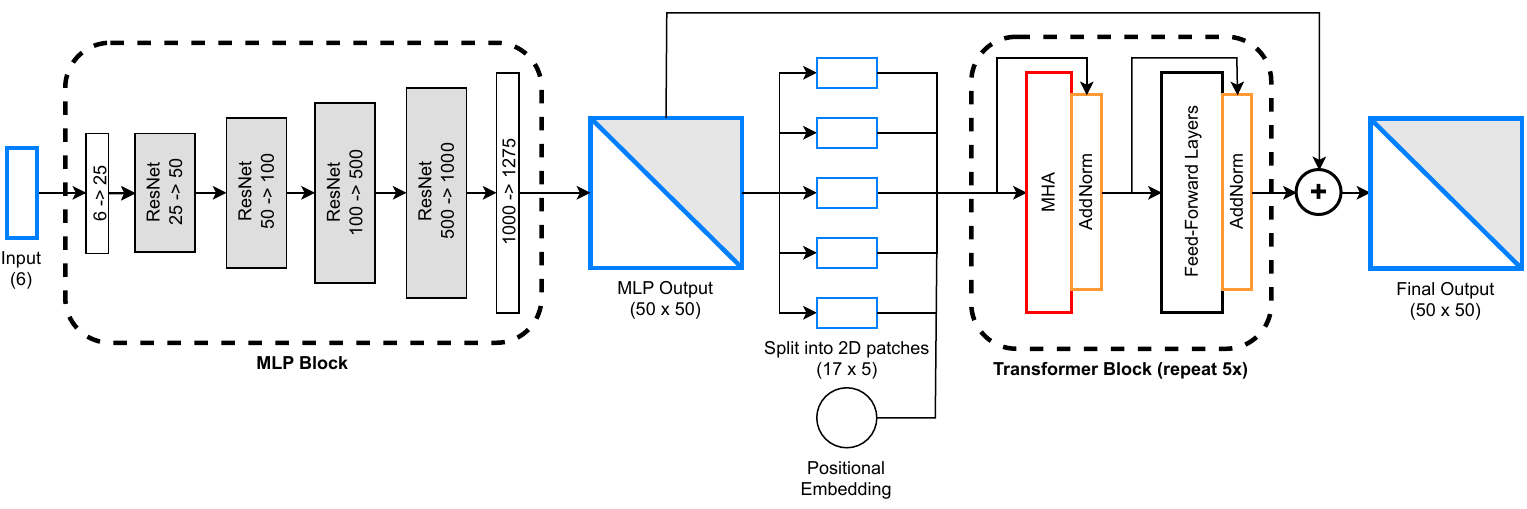}
    \caption{Neural network architecture of the covariance emulator. We split up the architecture into an MLP block and a transformer block, indicated by the dashed boxes. Cosmology parameters are first fed into a series of ResNet sub-blocks that each increases dimensionality. The output of these layers is then split into spatially-adjacent 2D patches. After positional embedding is applied, these patches are passed to a series of five transformer encoder blocks based on Ref. \protect\cite{Attention-Paper}. The output of these transformer blocks is added to the MLP block output. In this scheme, the transformer block is interpreted as applying a correction to the output from the MLP block.}
    \label{fig:MLP-T-chart}
\end{figure*}

Our emulator is composed of two main components: a multi-layer perceptron (MLP) block and a transformer block, as illustrated in Fig. \ref{fig:MLP-T-chart}. For a detailed background on machine learning with deep neural networks, we refer readers to Ref. \cite{D2L}.

The MLP block takes cosmology parameters as input and generates a lower-triangular matrix $L$ as output. The network begins with a fully connected layer to abstract the input cosmology into a feature vector. This vector is then passed through four residual network (ResNet) sub-blocks, each with an increasing number of nodes, enabling hierarchical complexity construction. Each ResNet sub-block consists of four fully connected layers with batch normalization, and ends with a residual connection with the input. The inclusion of batch normalization and residual connections serves to regularize network parameters, improve convergence, and facilitate information flow in deep networks \cite{BatchNorm, ResNet}. Finally, the encoded vectors from the ResNet sub-blocks are fed into another fully connected layer, transforming them into the targeted output dimension of $L$. Overall, the MLP block has eighteen MLP layers, twelve batchnorm layers, and four skip connections, totaling $6197900$ trainable parameters.

Next, we employ the transformer block as a higher-level correction to enhance the MLP's output. The transformer design, known for its outstanding performance in large language models like generative pre-trained transformers \cite{GPT4}, has also demonstrated high effectiveness in learning both tabular and image data \cite{FT_Transformer, ViT}.  Through its self-attention mechanism \cite{Attention-Paper}, transformers excel at capturing complex dependencies within input sequences, spanning both long and short ranges. This capability makes it particularly well-suited for learning the cosmological dependencies of covariance matrices or data vectors across scales ranging from large to small.

Our transformer block design follows a similar architecture to the encoder part of the vision transformer \cite{ViT}, which is specifically tailored for processing 2D structural data like images. The process starts with splitting the output from the MLP block into spatially-adjacent two-dimensional (2D) patches, which are then flattened into 1D sequences. We next employ positional embedding, which adds a location-dependent array to each patch such that individual elements retain information about their initial matrix position. This input then passes through five sequential transformer encoder sub-blocks, each with a multi-headed self-attention layer and a series of fully connected feed-forward layers. Both sub-block components are surrounded by an AddNorm connection, where the layer's input is added to its output and then normalized such that entries sum to one. The final output is then reorganized back into a 2D matrix and added to the MLP block output. In this sense, the transformer block attempts to apply a correction to the MLP output by extracting and interpreting the relationship between elements of the covariance matrix. Overall, the transformer block has $446335$ trainable parameters.

We have implemented our emulator using the \verb|PyTorch| python package \cite{PyTorch}. The specific sizes and dimensions of each layer in our network are detailed in Fig. \ref{fig:MLP-T-chart}. These network structural parameters are treated as hyperparameters and are established based on validation performance during training. For future applications, our package offers flexibility in adjusting layer dimensions and sub-layer configurations to accommodate varying input cosmology parameter sizes and covariance matrix dimensions.

\subsection{Neural Network Training}
\label{subsec:training-set}

We utilize mini-batch stochastic gradient descent to learn the network parameters that minimize the loss function given by

\begin{equation}
    l_n = \sum_{i,j} \left| \log{L}_{i,j}^\text{CovaPT} - \log{L}_{i,j}^\text{Emulator}\right|\,,
    \label{eq:loss}
\end{equation}
where $\log{L}^\text{CovaPT}$ is a Cholesky-decomposed covariance matrix calculated analytically, and $\log{L}^\text{Emulator}$ is the output from the neural network using the same input parameters. To minimize the loss function, we use the Adam optimization algorithm with a step function decaying learning rate \cite{Adam-optimization}. We use $80\%$ of the entire dataset for training purposes. The remaining $20\%$ is divided evenly into two sets: a validation set for adjusting the model's hyperparameters during training, and a test set for accurately evaluating the final model performance. 

\begin{figure}[h]
    \centering
    \includegraphics[width=\linewidth]{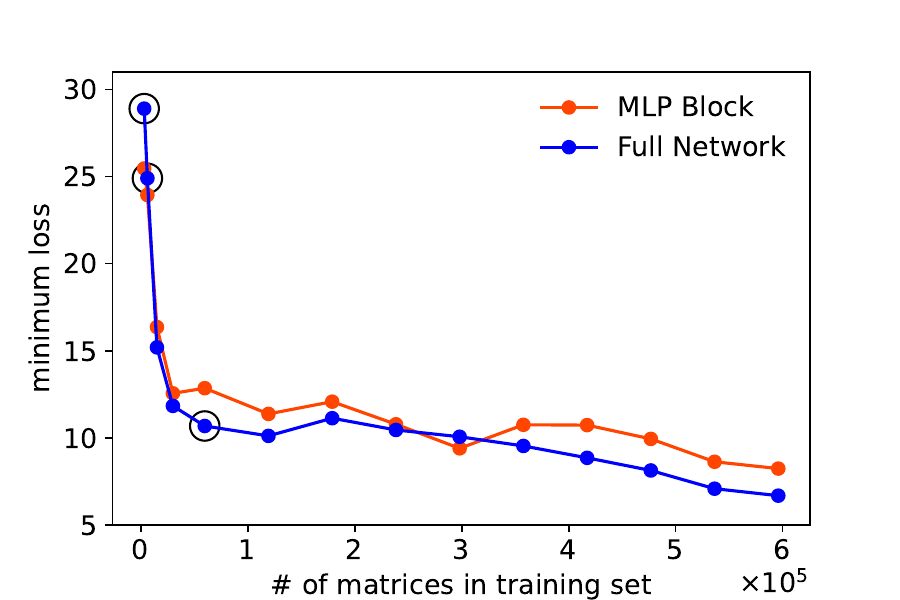}
    \caption{Minimum loss achieved for the MLP block (orange) and the full network (blue) with respect to training set size. Each point is the minimum of 2 independent training runs. Aside from the downward trend as the training size increases, it is clear the full network consistently performs better than the MLP network. Black circles correspond to networks used to run likelihood analyses in Fig. \ref{fig:Datasize-Test}.}
    \label{fig:loss-datasize}
\end{figure}

We train each block in our network separately, starting with the MLP block. After this phase is complete, we freeze the resulting MLP weights and proceed to train the transformer block using the same loss function. To prevent underfitting, we divide each training phase into three rounds with progressively decreasing initial learning rates. This approach allows later rounds to compensate for any suboptimal initial learning rates in earlier rounds, preventing the network from becoming trapped in local minima. To prevent overfitting, we implement an early stopping criterion, where training stops if the average loss on the validation set fails to improve for several epochs. This technique ensures that the network remains generalizable, while also cutting down on unnecessary training time.

Figure \ref{fig:loss-datasize} shows the results from training our emulator with varying fractions of the full training set. For each point shown, we take the best of two training rounds, with the minimum loss being the average value of Eq. \eqref{eq:loss} for the full test set. We can clearly see that the minimum loss value decreases quickly up to a training size of around $10\%$ of the full dataset, after which improvements begin to diminish. This trend suggests our network is not at the theoretical performance limit, since the loss is still improving with larger training sets. However, as we will discuss in the next section, we achieve satisfactory performance when using our complete training set w.r.t. calculating $\chi^2$ error and running simulated likelihood analyses.

\section{Emulator Performance Tests and Simulated Analyses}
\label{sec:tests}

\subsection{Chi Squared Test}

We calculate the $\chi^2$ for all matrices in the testing set via the following equation,

\begin{equation}
    \chi_i^2 = \left( P^{\text{data}} - P^{\text{bestfit}}\right)^T C(\theta)_i^{-1} \left( P^{\text{data}} - P^{\text{bestfit}}\right)\,,
\label{eq:chi-squared}
\end{equation}
where $P^{\text{data}}$ are the power spectrum multipoles presented in Ref. \cite{Beutler-Data},\footnote{\url{https://fbeutler.github.io/hub/deconv_paper.html}} and $P^{\text{bestfit}}$ is the model vector evaluated at the best-fit cosmology from an analysis of $P^\text{data}$ using a fixed analytic covariance matrix. Only $C(\theta)$ is varied as part of this test, and we calculate the difference between $\chi^2$ using an emulated covariance matrix vs a matrix in the test set to find the error.

\begin{figure*}
    \centering
    \includegraphics[width=0.81\linewidth]{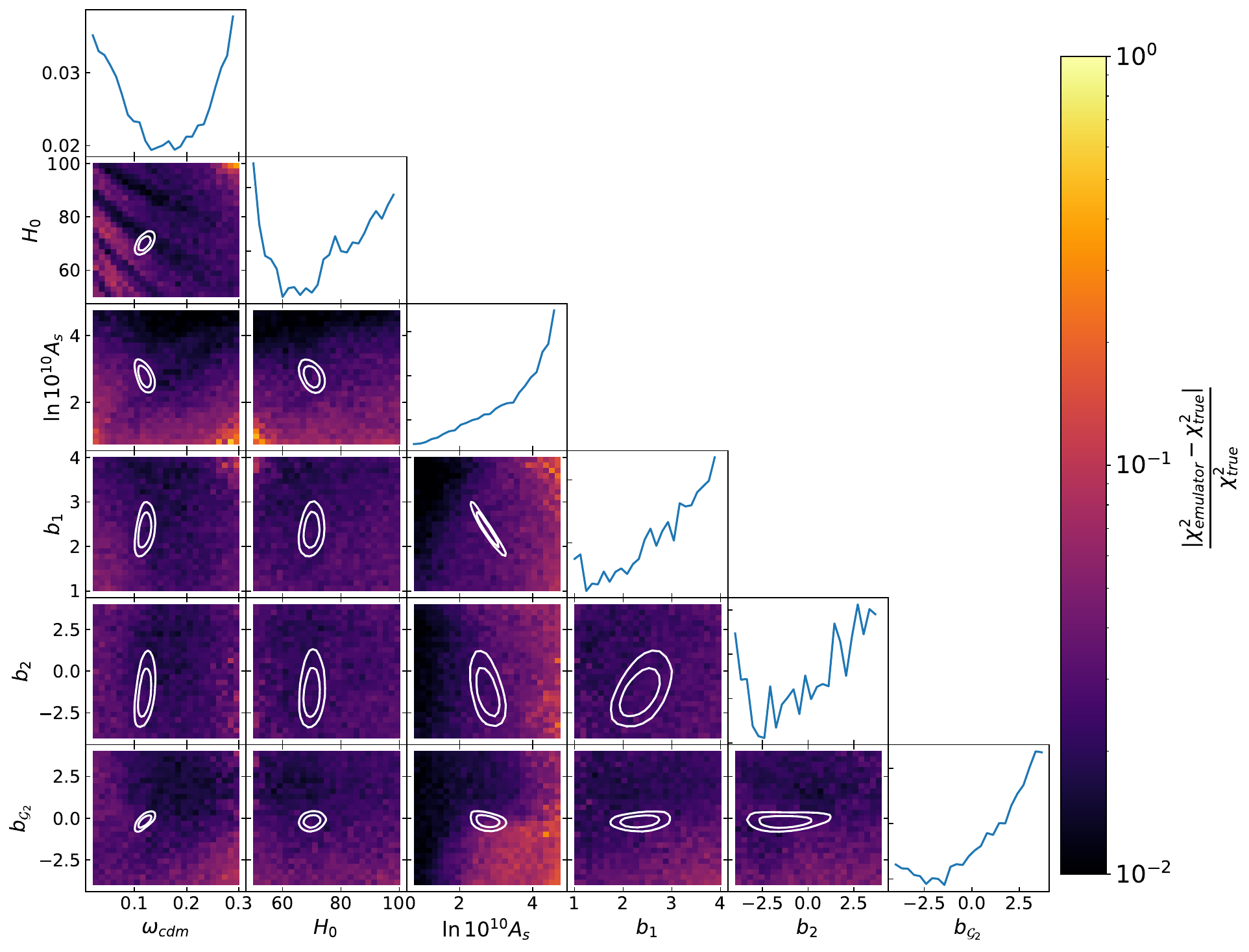}
    \caption{Corner heat map showing the relative $\chi^2$ error defined by Eq. \ref{eq:chi-squared} from emulating the full covariance matrix. Each pixel represents the mean $\chi^2$ error of matrices with parameter values within that pixel. 1$\sigma$ and 2$\sigma$ contours of a typical BOSS analysis are over-plot for reference. The $\chi^2$ error has some parameter dependence, particularly w.r.t. $\ln( 10^{10} A_s)$. However, we find this variance has no effect on our likelihood analysis tests.}
    \label{fig:chi2-error}        
\end{figure*}

Figure \ref{fig:chi2-error} shows the relative $\chi^2$ error over the entire emulated parameter space. We can see that the mean errors range from $~1\%$ to $~20\%$ with a slight parameter dependence in some of the panels. Globally, the $\chi^2$ error roughly follows a log-normal distribution with a mean value of $2.50\%$ and a $1 \sigma$ confidence region of $[0.465\%, 13.4\%]$. The parameter-dependence we observe could result in a shift in contours towards regions with smaller error during a likelihood analysis. However, our subsequent likelihood analysis tests do not exhibit this behavior, indicating that said dependence is small enough to be negligible.

\subsection{Simulated Analysis Details and Results}
\label{subsec:analysis}

\begin{table}[b]
\centering
    \begin{tabular}{cccc}
    \hline
    \hline
        Parameter & Fiducial value & Prior & Emulator Range\\ [0.5ex] 
    \hline
        $H_0$ [km/s/Mpc] & 67.77 & [50, 100] & Same as prior\\
        $\omega_{cdm}$ & 0.1184 & [0.02, 0.3] & Same as prior\\
        $\ln 10^{10} A_s$ & 3.0447 & [0.75, 4.75] & Same as prior\\
        $b_1$ & 2 & [1, 4] & Same as prior\\
        $b_2$ & 0 & $\mathcal{N}(0, 1)$ & [-4, 4]\\
        $b_{\mathcal{G}_2}$ & 0 & $\mathcal{N}(0, 1)$ & [-4, 4]\\
    \hline
        $c_0$ $[\text{Mpc/h}]^2$ & 0 & $\mathcal{N}(0, 30^2)$ & Marginalized\\
        $c_2$ $[\text{Mpc/h}]^2$ & 0 & $\mathcal{N}(0, 30^2)$ & Marginalized\\
        $\tilde{c}$ $[\text{Mpc/h}]^4$ & 500 & $\mathcal{N}(500, 500^2)$ & Marginalized\\
        $P_\text{shot}$ $[\text{Mpc/h}]^3$ & 0 & $\mathcal{N}(0, 5000^2)$ & Marginalized\\
    \hline
    \hline
    \end{tabular}
    \caption{Fiducial cosmology, priors, and emulator ranges used in our likelihood analyses. $[\text{min}, \text{max}]$ refers to a flat range, while $\mathcal{N}(\mu, \sigma^2)$ refers to a Gaussian distribution with mean $\mu$ and variance $\sigma^2$. Priors for cosmology parameters are set to be reasonably wide, while priors for nuisance parameters are taken from Ref. \protect\cite{Wadekar-2020}. Fiducial values are taken from Ref. \protect\cite{Planck-2018} or set to the center of their respective priors. The emulator range for a given parameter is set to the prior if it is flat, or cut off at $\pm 4 \sigma$ away from the center if it is Gaussian.}
    \vspace{1pt}
    \label{table:cosmology}
\end{table}

We follow the analysis choices of Ref. \cite{Wadekar-2020} and vary three cosmology parameters: the Hubble constant $H_0$, the physical dark matter density $\omega_{cdm}$, and the primordial power spectrum amplitude $A_s$. Both the spectral index and physical baryon density are fixed to their Planck 2018 best-fit values  ($n_s = 0.9649, \omega_b = 0.02237$) \cite{Planck-2018}, as previous analyses showed varying these parameters has little effect on the resulting contours \cite{BOSS, Wadekar-2020}. We also assume a single massive neutrino state with a fixed mass of $m_\nu = 0.06\ \text{eV}$.

Priors for both cosmology and nuisance parameters are given in Table \ref{table:cosmology}. Our emulator requires an explicit range in parameter space to work, hence the flat priors for ($H_0, \omega_{cdm}, A_s$). We sample the same nuisance parameters as described in Section \ref{subsec:Pk}. with priors taken from Ref. \cite{Wadekar-2020}. During our analyses, we analytically marginalize over $c_0$, $c_2$, $\tilde{c}$, and $P_\text{shot}$ using the procedure outlined in Ref. \cite{Analytic-marginalization}. Overall, our chains sample a total of six parameters, those being $[H_0, \omega_{cdm}, A_s, b_1, b_2, b_{\mathcal{G}2}]$. We present the resulting contours in terms of $[H_0, \Omega_m, \sigma_8]$ as those constraints are more prevalent in the literature. Additionally, we compute the best fit cosmology by calculating the mean parameter values of our output chain.

We assume a Gaussian likelihood for all our analyses with the form

\begin{equation}
    -2\ln \mathcal{L} = \ln|C| + \left[ \left( P^{\text{d}} - P^{\text{m}}\right)^T C^{-1} \left( P^{\text{d}} - P^{\text{m}}\right) \right]\,.
\label{eq:likelihood}
\end{equation}

\begin{figure}[h]
    \centering
    \includegraphics[width=\linewidth]{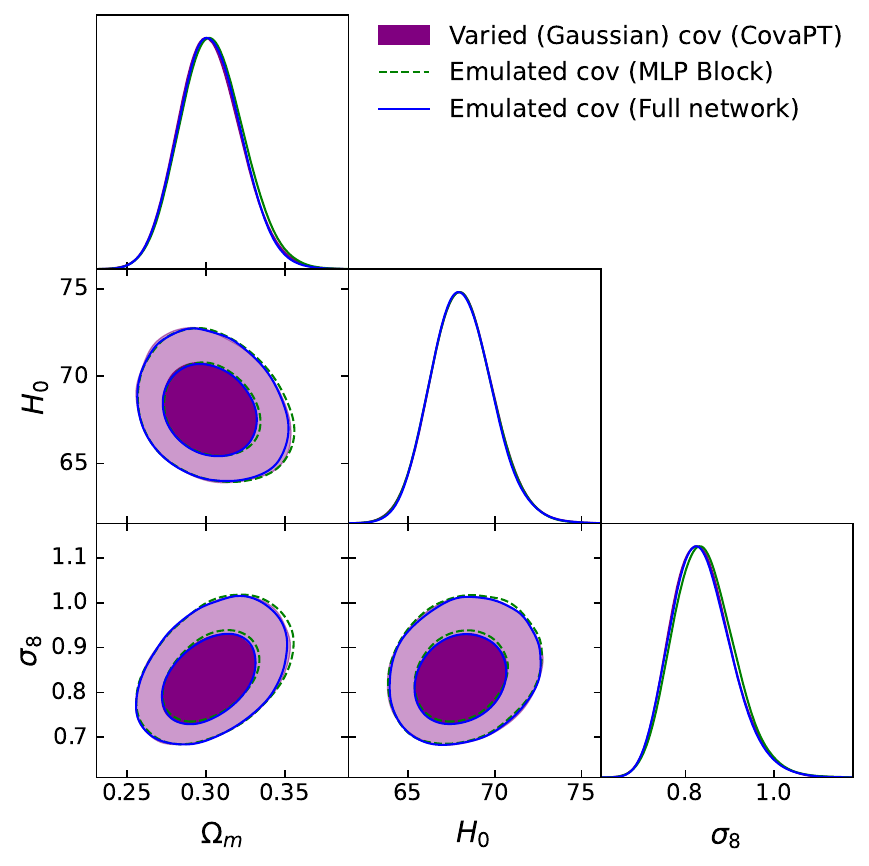}
    \caption{Parameter contours from simulated analyses with cosmology-dependent Gaussian covariance calculated via \texttt{CovaPT} (purple), the full emulator (blue), and the MLP block only (green). Chains were run with a simulated data vector and did not include the determinant term in the likelihood. The change in contours is small for both emulator configurations, and are most likely dominated by sampling noise.}
    \label{fig:gaussian-test}
\end{figure}

\begin{figure}
    \centering
    \includegraphics[width=\linewidth]{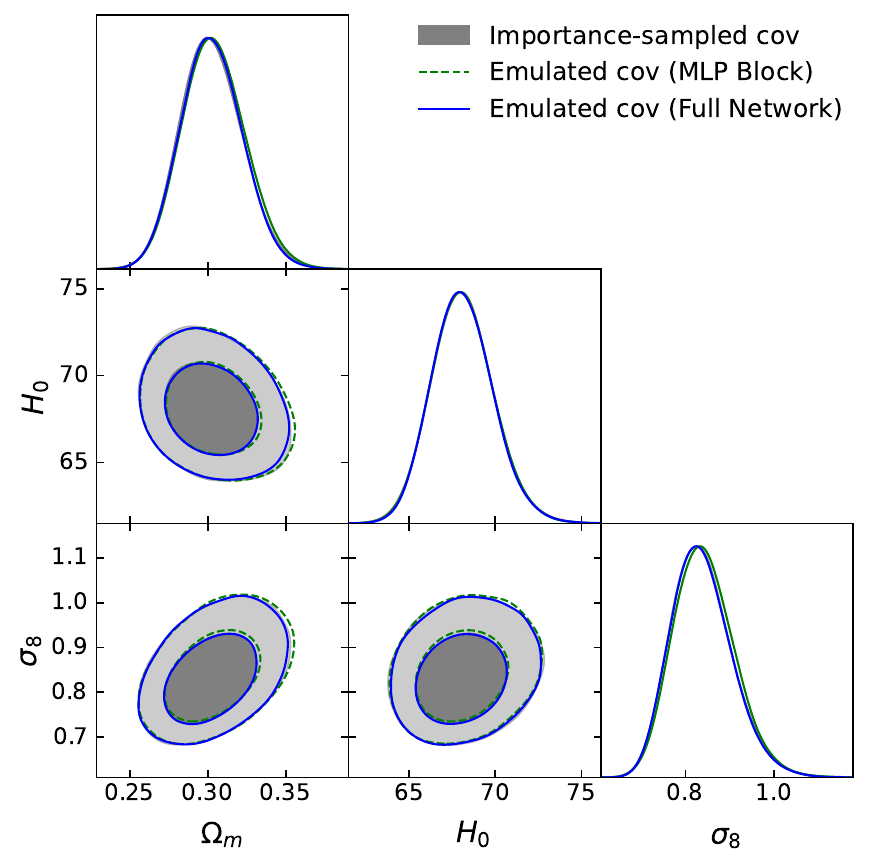}
    \caption{Parameter contours from simulated analyses with cosmology-dependent full covariance calculated with the full emulator (blue), and the MLP block only (green). Chains were run with a simulated data vector and without the determinant term in the likelihood. The gray reference contours were calculated using importance sampling, as running a chain with a full varied covariance matrix is too computationally expensive. Similar to Fig. \ref{fig:gaussian-test}, the change in contours when using both emulator configurations is negligible.}
    \label{fig:non-gaussian-test}
\end{figure}

In the simulated analyses that follow in Figs. \ref{fig:gaussian-test}, \ref{fig:non-gaussian-test}, and \ref{fig:Datasize-Test}, we neglect the determinant term and only vary the covariance in the exponential term. The reason is that we are only interested in studying the precision of our emulator, not in conducting a realistic analysis. For the BOSS data analysis in Fig. \ref{fig:BOSS-real} we show contours for both cases, with and without varying the covariance in the determinant. 

Our likelihood analyses were carried out using the \verb|MontePython| MCMC package \cite{Montepython}. We use the Gelman-Rubin criteria of $R < 0.01$ to determine that our chains are converged \cite{Gelman-Rubin, Brooks-Gelman}, and generate our contour plots with the \verb|GetDist| package \cite{GetDist}.

We first present the results of using running simulated analyses while emulating only the Gaussian term of the covariance. In this case, \verb|CovaPT| is sufficiently fast to be varied across the parameter space, so we can directly calculate a``truth'' contour and then compare to our emulator-based analysis. 

Figure \ref{fig:gaussian-test} compares results from using \verb|CovaPT| and our neural network emulator to generate Gaussian covariances. We consider two different emulator configurations, one consisting of the MLP block only and another that uses the full network. For both network configurations, the resulting contours are almost directly on top of each other. We calculate the 1D shift in best-fit parameters to range from $[0.025 - 0.1]\sigma$ for the MLP block, and $[0.007 - 0.08]\sigma$ for the full emulator. The specific shift values for each parameter are given in Table \ref{table:loss-data}. We also find that the error bars for each parameter agree to within $2\%$ for the MLP block, and $1.7\%$ for the full emulator.

Next, we present our results from emulating the full covariance matrix at every point in an MCMC chain. For the comparison contours, we take the Gaussian covariance chain from the previous test and re-calculate the full covariance matrix (and subsequently the likelihood) for each set of parameters. The chain is then re-weighed in a process similar to importance sampling. This process is necessary because full covariance matrices are too expensive to calculate during an MCMC analysis.

Figure \ref{fig:non-gaussian-test} shows the posterior distributions from our importance-sampling method, and from using our emulator to vary the full covariance matrix. Similar to the previous test, the contours lie almost directly on top of each other, indicating our emulator functions at high accuracy. We again calculate shifts in best-fit values to range from $[0.006 - 0.09] \sigma$ for the MLP block, and $[0.015 - 0.15]\sigma$ for the full emulator. The specific shift values for each parameter are given in Table \ref{table:loss-data}. We also find that the error bars for each parameter agree to within $5\%$ for the MLP block, and $4.6\%$ for the full emulator.

Our emulator can reproduce contours slightly better when emulating Gaussian instead of non-Gaussian covariance matrices, which is likely due to the fact that Gaussian covariances have fewer off-diagonal components. The loss values reported in Table \ref{table:loss-data} corroborate this hypothesis, as the reported minimum loss is much smaller when emulating the Gaussian term only. However, it is important to keep in mind that part of this increased uncertainty for the non-Gaussian covariance can also be caused by the importance sampling technique. Unfortunately, there is no perfect truth contour to compare to in the case of non-Gaussian covariances. Keeping these effects in mind for the latter case, our emulator is able to reproduce the expected contours with differences small enough to be statistically insignificant.

\begin{table*}[t]
    \begin{ruledtabular}
    \begin{tabular}{ccccc}
    \multicolumn{1}{c}{\textrm{Network}}&
    \multicolumn{1}{c}{\textrm{MLP Block (Gaussian)}}&
    \multicolumn{1}{c}{\textrm{Full (Gaussian)}}&
    \multicolumn{1}{c}{\textrm{MLP Block}}&
    \multicolumn{1}{c}{\textrm{Full}}\\
    \colrule
    Loss & 0.620 & 0.452 & 8.24 & 6.68 \\
    \hline
    $H_0$ & 0.007$\sigma$ & 0.007$\sigma$ & 0.006$\sigma$ & 0.029$\sigma$ \\
    $\omega_{cdm}$ & 0.065$\sigma$ & 0.003$\sigma$ & 0.078$\sigma$ & 0.086$\sigma$ \\
    $\ln 10^{10} A_s$ & 0.041$\sigma$ & 0.009$\sigma$ & 0.008$\sigma$ & 0.015$\sigma$\\
    $b_1$ & 0.042$\sigma$ & 0.013$\sigma$ & 0.007$\sigma$ & 0.026$\sigma$ \\
    $b_2$ & 0.057$\sigma$ & 0.026$\sigma$ & 0.087$\sigma$ & 0.151$\sigma$ \\
    $b_{\mathcal{G}_2}$ & 0.060$\sigma$ & 0.026$\sigma$ & 0.072$\sigma$ & 0.093$\sigma$ \\
    \hline
    $\Omega_m$ & 0.078$\sigma$ & 0.007$\sigma$ & 0.085$\sigma$ & 0.121$\sigma$\\
    $\sigma_8$ & 0.082$\sigma$ & 0.007$\sigma$ & 0.064$\sigma$ & 0.042$\sigma$  \\
    \end{tabular}
    \caption{Observed absolute shift in 1D best-fit values when comparing chains made with our covariance emulator with those made with analytical calculations. The 2nd and 3rd columns correspond to the chains in Fig. \ref{fig:gaussian-test}, while the 4th and 5th columns correspond to Fig. \ref{fig:non-gaussian-test}. The upper rows show parameters that were sampled directly, while the bottom two rows show our derived parameters. Almost all parameters show a comparable shift for the full emulator compared to the MLP block. We also show the average loss achieved by each network for the test set.}
    \label{table:loss-data}
    \end{ruledtabular}
\end{table*}

We also see marginal differences between using the full emulator and MLP block individually. There is a small improvement in matrix accuracy when using the full emulator, as indicated by the loss values in Table \ref{table:loss-data}. We also see a small improvement in error bars for both the Gaussian and full covariance contours. However, for the full covariance case, the full emulator has slightly larger shifts in best-fit values than the MLP block, which is counter to the behavior for Gaussian covariance. These results suggest that the transformer block has a small effect on performance in practice.

\subsection{Effect of Training Set Size}

\begin{figure}
    \centering
    \includegraphics[width=\linewidth]{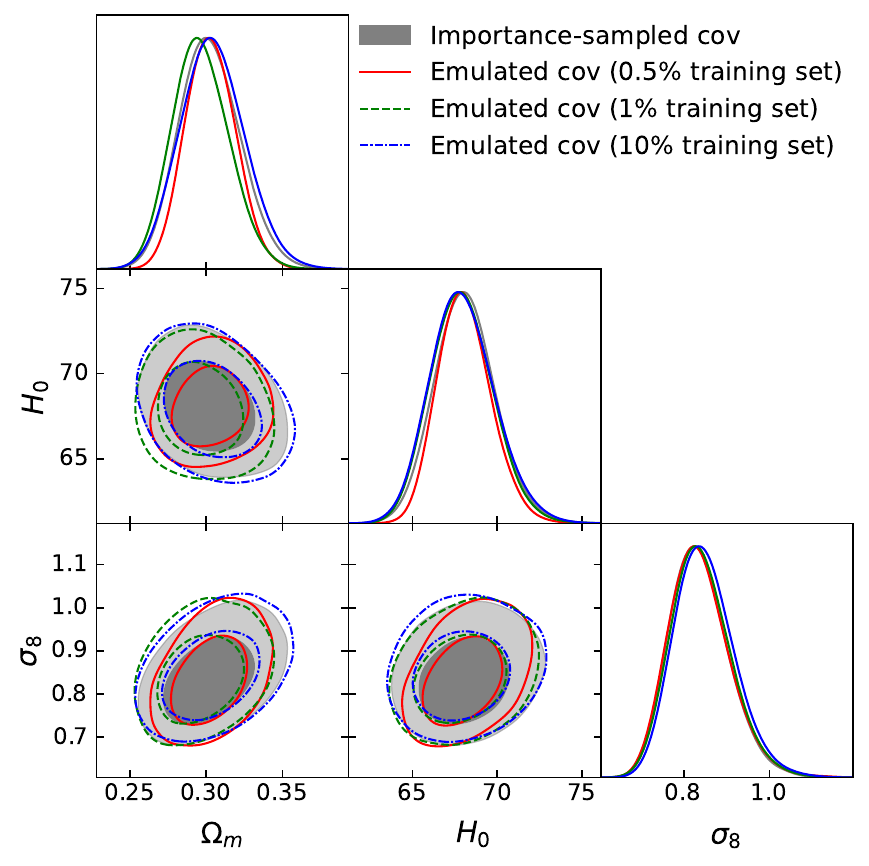}
    \caption{Same as Fig. \ref{fig:non-gaussian-test}, except the emulators are trained on $0.5\%$ (red), $1\%$ (green), and $10\%$ (blue) of the full training set. As the number of training samples increase, the emulator contours get closer to the``tru'' gray contours. The blue contours approach the same level of deviations as the full network, which suggests $6 \times 10^4$ matrices is the required training set size for an accurate emulator w.r.t. a likelihood analysis.}
    \label{fig:Datasize-Test}
    \vspace{-10pt}
\end{figure}

To better quantify how the size of the training set affects the accuracy of our emulator, we run likelihood analyzes with networks trained on $0.5\%$, $1\%$, and $10\%$ of our full training set. These networks are indicated by the black circles in Fig. \ref{fig:loss-datasize}, and the resulting contours are presented in Fig. \ref{fig:Datasize-Test}. Visually, the contours deviate significantly from the ``truth'' contours for the $0.5 \%$ and $1 \%$ emulators, with the emulator trained on $10 \%$ of the training set approaching that of the full network presented in Fig. \ref{fig:non-gaussian-test}. We find the range in best-fit value shifts, in order of smallest training size to largest, are $[0.04 - 0.08] \sigma$, $[0.06 - 0.33] \sigma$, and $[0.02 - 0.18]\sigma$. These results indicate that our emulator achieves an effective performance when trained on at least $6 \times 10^4$ covariance matrices.

\section{\label{sec:BOSS}BOSS re-analysis}
In this section, we use our covariance matrix emulator to perform a re-analysis of BOSS DR12 galaxy power spectrum data. We use the power spectrum monopole and quadropole presented in Ref. \cite{Beutler-Data} up to $k_\text{max} = 0.25$ h/Mpc as our data vector, and compute the model vector as described in Section \ref{subsec:Pk} using the public code \verb|CLASS-PT| \cite{CLASS-PT}.

\subsection{BOSS DR12 Data}

We perform three different analyses with the BOSS DR12 data:

\begin{table*}[]
\renewcommand{\arraystretch}{1.4}
    \begin{ruledtabular}
    \begin{tabular}{c|cc|ccc}
    \textrm{Parameter}& \textrm{This work (Analysis 1)}& \textrm{This work (Analysis 3)} & Halo model \cite{Kobayashi22}& EFTofLSS \cite{Philcox_Ivanov22} & LPT \cite{Chen22}\\
    & (HighZ NGC) & (HighZ NGC) & (HighZ NGC) & (Full BOSS) & (Full BOSS)\\
    \hline
    $\Omega_m$ & $0.293\pm 0.017$ & $0.276^{+0.013}_{-0.015}$ & $0.298^{+0.017}_{-0.020}$ & $0.312^{+0.011}_{-0.012}$&  $0.305 \pm 0.01$\\ 
    $H_0$  & $70.3\pm 2.0$ & $70.2\pm 1.9$ & $70.3^{+2.2}_{-2.1}$ & $68.5^{+1.1}_{-1.3}$ & $68.5\pm 1.1$  \\
    $\sigma_8$  & $0.702^{+0.063}_{-0.075}$ & $0.674^{+0.058}_{-0.077}$ & $0.685^{+0.074}_{-0.044}$ & $0.737^{+0.040}_{-0.044}$&  $0.738 \pm 0.048$ \\
    \end{tabular}
    \end{ruledtabular}
    \caption{Mean values and $68\%$ minimal credible intervals of the green (fixed covariance) and red (varied covariance) parameter contours presented in Fig. \ref{fig:BOSS-real}. We also show results from the full-shape analyses done by Refs. \cite{Chen22, Kobayashi22, Philcox_Ivanov22} which use the same power spectrum data, but all employ different analysis methods and scale cuts. From Ref \cite{Kobayashi22}, we only extract constraints from the high redshift, NGC sample for better comparison with our results. Refs \cite{Chen22, Philcox_Ivanov22} only report results from analyzing the full BOSS dataset, which explains their improved constraining power compared to analysis 1. $H_0$ is given in units [km s$^{-1}$ Mpc$^{-1}$].}
    \label{tab:BOSS-analysis}
\end{table*}

\begin{figure}
    \centering
    \includegraphics[width=\linewidth]{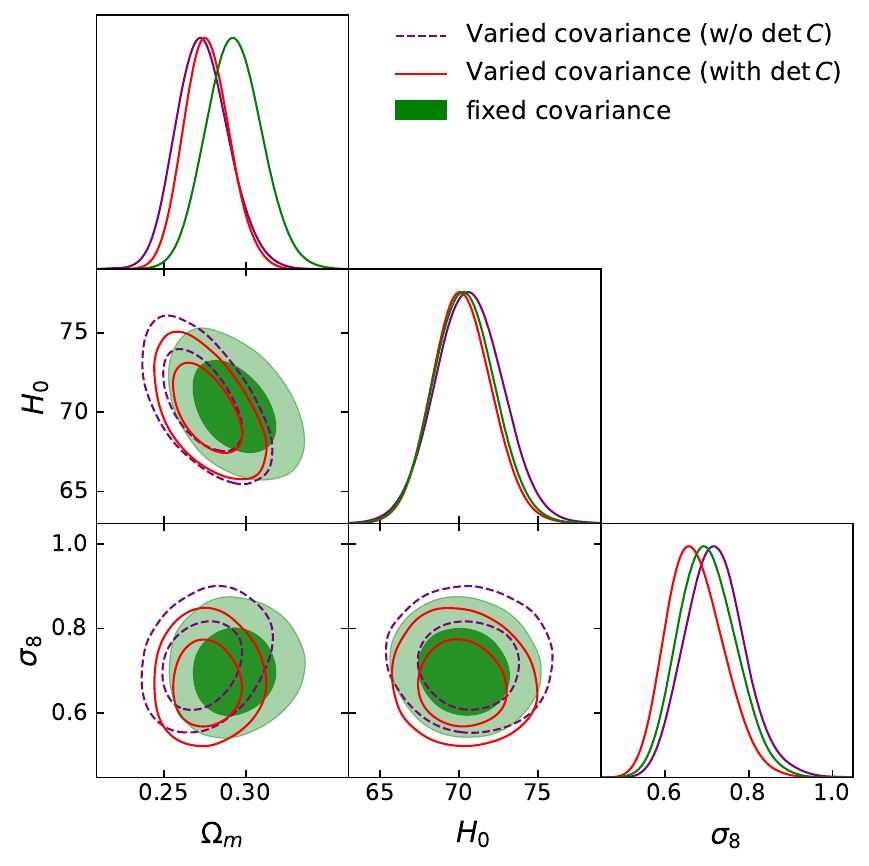}
    \caption{Parameter contours from analyzing the high redshift, NGC chunk of BOSS DR12 data generated with a fixed covariance matrix (green), varied covariance without the determinant term in the likelihood (purple), and varied covariance with the determinant term (red). Varying the covariance with cosmology results in an overall shift in contours and a slight reduction in error bars, with small differences when including the determinant term in the likelihood.}
    \label{fig:BOSS-real}
\end{figure}

\begin{itemize}
    \item Analysis 1: We calculate the covariance matrix using \verb|CovaPT| at a fixed cosmology and keep it fixed throughout the MCMC analysis. Subsequent chains are run with an updated input cosmology for the covariance until the output has converged.
    \item Analysis 2: We calculate the full covariance using our emulator at every point of the chain, and neglect the determinant term in the likelihood given by Eq. \ref{eq:likelihood}.
    \item Analysis 3: We again calculate the full covariance at every point of the chain, this time including the determinant term in the likelihood.
\end{itemize}

The resulting contours are given in Fig. \ref{fig:BOSS-real}, and the 1D marginalized constraints for analysis 1 and analysis 3 are given in Table \ref{tab:BOSS-analysis}. The chains from analysis 1 have an average $0.93 \sigma$ offset compared to previous BOSS analyses (e.g. Refs. \cite{Ivanov-2020, Wadekar-2020}), which is expected since we are using the updated BOSS DR 12 data vector from Ref. \cite{Beutler-Data}. 

We also present results from BOSS analyses of the same data vector using EFT \cite{Philcox_Ivanov22}, LPT \cite{Chen22}, and halo model based \cite{Kobayashi22} theoretical models. When comparing analysis 1 to the results of Refs. \cite{Philcox_Ivanov22, Chen22} , our values are slightly shifted and have larger error bars than expected from using different models and scale cuts (e.g. $\Omega_m = 0.293\pm 0.017$ has larger error bars than EFT and LPT results by a factor of $\sim1.7$). These differences are most likely due to those studies analyzing the full BOSS dataset, while we restrict ourselves to the high redshift NGC data sample. We see much better agreement when comparing our results to the ones in Ref. \cite{Kobayashi22}, who also give results for the same data sample.

When comparing analysis 1 and analysis 2, we find an average shift of $0.71 \sigma$ in the 1D best fit values and a $10\%$ reduction in the error bars. As a specific example, $\Omega_m$ is shifted down by $1.12 \sigma$ from $0.293$ to $0.274$, Similarly, we see an average $0.46\sigma$ shift and a $5\%$ improvement in constraining power when comparing analysis 1 to analysis 3 ($\Omega_m$ is shifted down by $0.97 \sigma$ to a value of $0.276$). These results suggest that cosmology dependent covariance impacts the resulting parameter contours, and said impact is partially due to the determinant term in Eq. \ref{eq:likelihood}.

\subsection{Comparison with simulated analyses}

We perform several simulated analyses with a cosmology-dependent covariance matrix in order to explore whether the shifts in best fit values and changes in contour size we observe between analysis 1 and analysis 3 are expected. These simulated analyses use noisy data vectors, where the noise is drawn from a multivariate Gaussian distribution whose width is defined by the covariance computed at the fiducial cosmology. 

To investigate the impact of individual realizations of our noisy data vector, we also perform an analysis maximizing the following likelihood function

\begin{equation}
    \mathcal{L}_\text{avg} = \frac{1}{N_s}\sum_i^{N_s} \mathcal{L}_i\,,
\end{equation}

where $N_s$ is the number of realizations, and $\mathcal{L}_i$ is the likelihood calculated using Eq. \ref{eq:likelihood} and an individual noisy simulated data vector. This average likelihood essentially marginalizes over individual realizations, and is similar to the analysis done in Ref. \cite{Euclid-Covariance}. We took the average of $N_s = 100$ realizations in this way.

\begin{figure*}
    \centering
    \begin{minipage}{.5\textwidth}
        \centering
        \includegraphics[width=\linewidth]{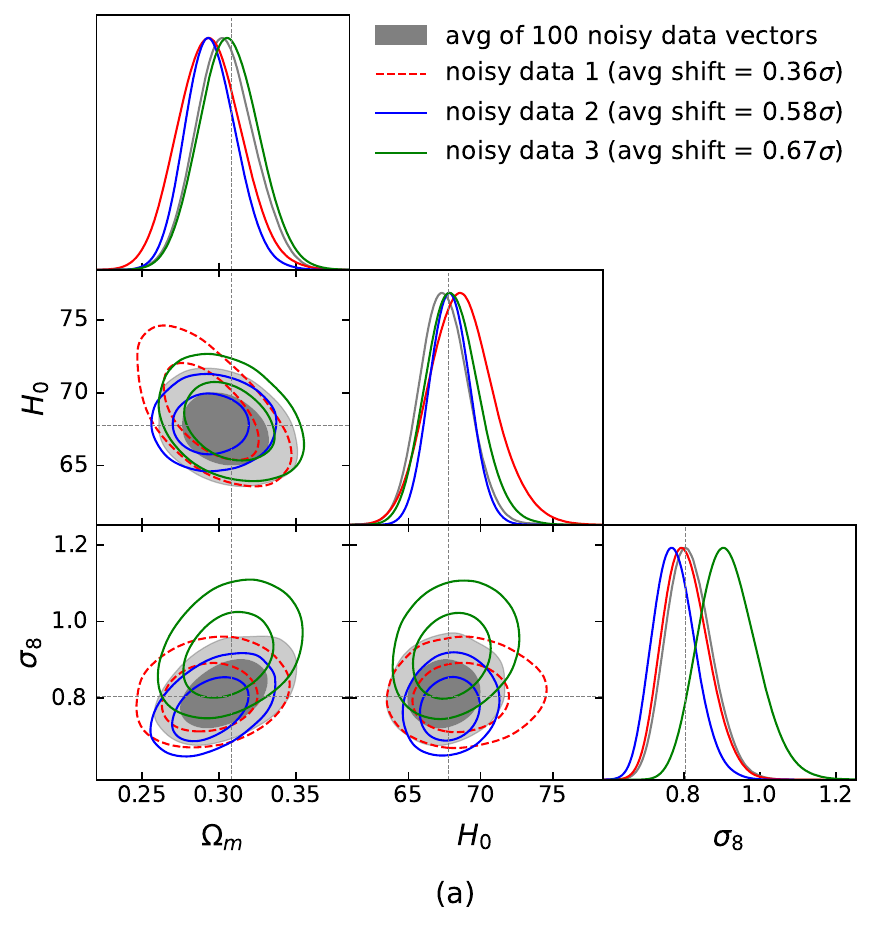}
    \end{minipage}%
    \begin{minipage}{.5\textwidth}
        \centering
        \includegraphics[width=\linewidth]{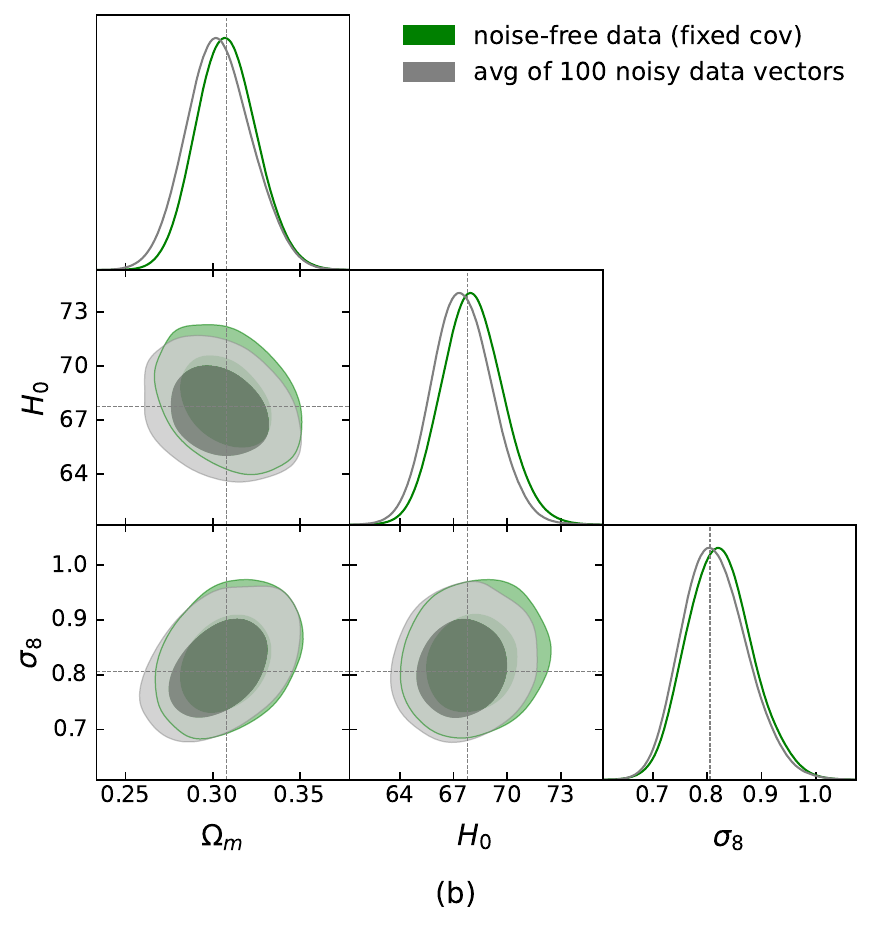}
    \end{minipage}
    \caption{(a). Parameter contours from simulated analyses using our emulator to vary covariance with cosmology. Simulated data vectors were drawn from a multivariate Gaussian defined by the fiducial model vector and covariance matrix. Several chains with individual realizations are shown, with the averaged likelihood constraints given in gray. We see a similar level of shift from the fiducial cosmology for individual data realizations as the red contours in Fig. \ref{fig:BOSS-real}.\\
    (b). Contours from our averaged likelihood chain (gray) compared to a chain run with noise free simulated data and a fixed covariance matrix (green). Both contours lie on top of each other, despite the differences in treatment of the covariance matrix.}
    \label{fig:BOSS-noisy-analysis}
\end{figure*}

Figure \hyperref[fig:BOSS-noisy-analysis]{8a} shows the results of running chains with several different data realizations, as well as our averaged likelihood chain. Individual realizations appear to result in different size contours and shifts from the fiducial cosmology. Taken as a whole, however, the best-fit shift is a similar size as the difference between varied and fixed covariance in the real BOSS analysis. This result suggests the shift we see in that analysis is within expectations.

We also see that the averaged likelihood chain is centered around the fiducial cosmology. Figure \hyperref[fig:BOSS-noisy-analysis]{8b} shows those contours in comparison with a simulated analysis run with a noise-free data vector and a fixed covariance matrix. These contours lie almost directly on top of each other, despite the different covariance matrix treatments.

\section{Conclusion}
\label{sec:conclusion}

The calculation of high precision covariance matrices for cosmological observables is computationally too expensive to be performed on the fly during an MCMC, even if they are modelled analytically. To combat this problem, we presented a novel method of generating covariance matrices using a combination MLP and transformer network architecture. We used this emulator to generate analytic covariance matrices for the high redshift, north galactic cap sample of the BOSS DR12 galaxy catalog. After training, our emulator outputs a covariance matrix in less than $10$ ms on a typical desktop CPU, roughly five orders of magnitude faster than an analytical covariance model. The code used to build this network has been made publicly available on GitHub.\footnote{\url{https://github.com/jadamo/CovNet}}

We tested the accuracy of the emulator by computing the relative $\chi^2$ error when using covariance matrices from our emulator vs analytic calculations, and by running simulated likelihood analyses with cosmology dependent Gaussian and full non-Gaussian covariances. We found that our emulator has an average $2.5 \%$ $\chi^2$ error, with a small but negligible input parameter dependence. For the Gaussian covariance, our emulator reproduced the expected contours to great precision, with best fit shifts less than $0.08 \sigma$ and error bar agreement to within $1.7 \%$). Likewise, varying the full non-Gaussian covariance resulted in slightly larger, but still insignificant, uncertainties with individual parameter shifts less than $0.15 \sigma$ and error bar agreement within $4.6 \%$.

When comparing emulator performances, we found that using the full emulator (transformer and MLP) provides marginal improvements compared to using the MLP block only. We also determined a training set size threshold of $6 \times 10^4$ matrices where network performance begins to plateau. This threshold can serve as an initial size when building an emulator for other cosmological probes, although the optimal amount will depend on the specific data vector and parameter space considered.

We then use our emulator to investigate the impact of cosmology-dependent covariances in the context of a BOSS DR12 re-analysis. When keeping the covariance matrix fixed, we found our parameter constraints ($\Omega_m = 0.293\pm 0.017$, $H_0 = 70.3\pm 2.0$ km/s/Mpc, $\sigma_8 = 0.702^{+0.063}_{-0.075}$) broadly agree with other studies of the HighZ NGC data sample. We also found that varying covariance with cosmology results in a notable $0.46\sigma$ average shift in best-fit and $5\%$ improvement in constraining power ($\Omega_m = 0.276^{+0.013}_{-0.015}$, $H_0 = 70.2\pm 1.9$ km/s/Mpc, $\sigma_8 = 0.674^{+0.058}_{-0.077}$) compared to keeping the covariance matrix fixed. To better understand this difference, we repeated our analysis with simulated noisy data vectors and found a similar level of shift from the fiducial cosmology. We then ran a chain maximizing the average likelihood of $100$ noisy data realizations, and found doing so broadly recovers the results from analyzing a noise-free data vector with a fixed covariance matrix.

This work demonstrates that more complex cosmological quantities than just two-point statistics (see e.g. Refs. \cite{Manrique-Yus-2020, COSMOPOWER, Saraivanov-2024, Boruah-2023}) can be emulated at high accuracy over a sufficiently larger parameter range to run state-of-the-art likelihood analysis. We consider the idea of training neural networks to emulate summary statistics and their covariances to be a promising way forward when it comes to running the many simulated likelihood analyses needed to optimize analysis choices of future surveys.  

\begin{acknowledgments}
We would like to thank Mikhail Ivanov for providing assistance with the likelihood code, and Yosuke Kobayashi and Annie Moore for helpful modelling discussions. This work is supported by the Department of Energy HEP-AI program grant DE-SC0023892 and the Cosmic Frontier program grant DE-SC0020215. We acknowledge support from the SPHEREx project under a contract from the NASA/GODDARD Space Flight Center to the California Institute of Technology. The emulator developed in this paper used High Performance Computing (HPC) resources supported by the University of Arizona TRIF, UITS, and RDI and maintained by the UA Research Technologies department. 
\end{acknowledgments}

\bibliographystyle{apsrev4-2}
\bibliography{sources}

\end{document}